\documentclass[%
 preprint, %linenumbers,
superscriptaddress,
 amsmath,amssymb,
 aps, %physrev,
 prfluids
]{revtex4-2}

\usepackage{graphicx}% Include figure files
\usepackage{dcolumn}% Align table columns on decimal point
\usepackage{bm}% bold math
\begin{document}

% \preprint{APS/123-QED}

\title{\textbf{Multi-pathline flow visualization using PIV images} 
}% 

\author{Yukun Sun}
\affiliation{Department of Biological and Environmental Engineering, Cornell University, Ithaca, NY 14853, USA}

\author{Elijah James}%
\affiliation{Department of Biological and Environmental Engineering, Cornell University, Ithaca, NY 14853, USA}

\author{Frank Fang}%
\affiliation{Department of Biological and Environmental Engineering, Cornell University, Ithaca, NY 14853, USA}

\author{Jasper Agrawal}%
\affiliation{Department of Biological and Environmental Engineering, Cornell University, Ithaca, NY 14853, USA}

\author{Christopher Dougherty}%
\affiliation{Department of Biological and Environmental Engineering, Cornell University, Ithaca, NY 14853, USA}

\author{Cong Wang}
\affiliation{Department of Mechanical Engineering, University of Iowa, Iowa City, IA 52242, USA}

\author{Chris Roh}
\email{cr296@cornell.edu}
\affiliation{Department of Biological and Environmental Engineering, Cornell University, Ithaca, NY 14853, USA}

% \collaboration{CLEO Collaboration}%\noaffiliation


\begin{abstract}
One of the oldest flow visualization techniques is through  multiple pathlines generated by the movement of seeding particles spatially distributed in the flow. In the computerized era, particle images are used in quantitative measurements, such as particle image and particle tracking velocimetry (PIV and PTV). Here, we present several methods for post-processing raw particle images to generate enhanced flow visualization without a need for conducting additional experiments. Three post-processing methods will be shown: 1) controlling the exposure time, 2) color-coding temporal information, and 3) changing the frame of reference. We showcase how employing these three methods can highlight different flow features in three canonical flow cases: vortex ring, leading edge vortex, and turbulent boundary layer. In addition to the quantitative flow field, the multi-pathline visualization is expected to augment our ability to observe fluid flow from many different perspectives.

% \begin{description}
% Secondary publications and information retrieval purposes.
% You may use the \texttt{description} environment to structure your abstract;
% use the optional argument of the \verb+\item+ command to give the category of each item. 
% \end{description}
\end{abstract}

\maketitle

%\tableofcontents

\section{Introduction}
Flow visualization qualitatively and quantitatively highlights the salient features of fluid flow. One of the oldest and most intuitive methods for visualizing a flow field is pathline visualization \cite{prandtl_tietjens_1934,CodexArundel162r,CodexWindsor172v}. A pathline is defined as the history of the path taken by a particle \cite{national1972illustrated}. By seeding the fluid with multiple particles and simultaneously exposing them to a camera for an extended time, each light-scattering particle leaves its footprint on the imaging plane, revealing the fluid flow. 

Each individual pathline is a Lagrangian description of the fluid defined as follows:
\begin{equation}
    \underline{x}(t) = \underline{x}_0 + \int_0^t \underline{u}(\underline{x},t')\,dt \qquad \text{for } t \in [0,T],
\end{equation}
where $\underline{x}$ is the coordinate of the particle location at time $t$, $\underline{x}_0$ is the initial particle location at $t=0$, and $\underline{u}$ is the velocity of the particle. Continuous collection of particle location from $t=0$ to $T$ gives the pathline.

Spatially distributed multiple short pathlines can show an instantaneous flow field, yielding an Eulerian view of the fluid. Such flow visualization is often referred to as particle streaks. However, this is not to be confused with a streakline, which has a strict definition in fluid mechanics as a line formed by the collection of particles released at the same location at different times. To avoid confusion, we will henceforth use multi-pathline flow visualization instead of particle streak visualization.

The visualization using multi-pathline dates back to Leonardo da Vinci, who seeded the flow with particulates such as panic grass seeds, meshed paper, sand, and bubbles \cite{annrev_leonardo,CodexArundel162r,CodexWindsor172v}. Other artists who depicted prominent features in art perhaps also gained insight into motion when the flow was opportunistically visualized through the presence of natural particles like leaves. This same method was extensively used by Ludwig Prandtl, whose textbook \cite{prandtl_tietjens_1934} almost exclusively depicts flow phenomena using multi-pathline visualization. In “An Album of Fluid Motion” \cite{vanDyke_1982}, many figures are illustrated using magnesium or aluminum particulates or air bubbles by Taneda \cite{taneda1956b}, Werle \cite{werle1962separation}, and others \cite{coutanceau1968}.

With the advent of quantitative flow measurement \cite{gharib_piv,dana_dabiri_ptv,annrev_ptv}, such as particle tracking velocimetry (PTV) and particle image velocimetry (PIV), the old art of multi-pathline visualization has taken a back seat. However, the multi-pathline visualization can complement quantitative methods by revealing the field of curved particle trajectories and providing an algorithmic-error-free baseline for comparison with PTV and PIV.

These benefits can be unlocked without additional experimentation, once the raw particle images are obtained for PIV. Simply stacking the raw images can create multi-pathline visualization. In addition to this simplicity, the highly temporally-resolved sequence of particle images also offers flexibility in how the images can be stacked. Some of the techniques that needed to be controlled experimentally in the past \cite{prandtl_tietjens_1934}, can now be controlled during the post-processing. This convenience allows rapid observation of the same flow from different perspectives, which can offer further insight and enhance aesthetics.

Here, we demonstrate several methods for post-processing raw particle images to generate multi-pathline visualizations. We show three post-processing methods that 1) control exposure time (pathline integration length), 2) encode temporal information in color, and 3) change the frame of reference. To demonstrate the perspective changes these methods bring about, methods applied to three cases are shown: 1) vortex ring formation, 2) leading edge vortex roll-up, and 3) turbulent boundary layers. %To demonstrate unsteady motion of the vortex, stopping vortex shed from rotating plate will be shown at the end.

\section{Flow generation and raw particle image acquisition}
In all three cases, the raw particle images are recorded with a high-speed camera in a laboratory frame and a continuous-wave laser sheet illuminating the particles. Details of the flow generation, equipment used, and camera settings are listed below.

\subsection{Vortex ring}\label{sec:setup vortex ring}
The vortex ring was generated from a bent straw as a circular cylinder. A plastic straw with diameter of 0.23 in was filled with water from an external reservoir after which it was transferred into a plexiglass still water tank that is 100 cm long, 100 cm wide, and 15 cm deep. Positioning the straw in the tank such that the water column height in the straw above that of the tank created a hydrostatic pressure head used to drive vortex ring formation.

The fluid in the tank and in the separate external reservoir were both seeded with silver-coated glass spheres (Potters Industries LLC). The particles were illuminated by a 1 mm thick laser sheet generated from 2W 532 nm continuous laser (Ultralasers Inc). The laser beam passed through a plano-concave lens and a bi-convex lens to generate the laser sheet, which was aligned with the center of the straw nozzle. The image sequences of the vortex ring were collected by a high-speed camera (Krontech Chronos 2.1) with a 100 mm f/2.8 macro lens (Venus Optics Laowa). The camera was consistently operated at 1000 frames per second with an exposure time of 250 $\mu$s.
 
\subsection{Impulsively started airfoil at $30^\circ$ AoA}
The leading edge vortex (LEV) was generated by a flat plate moving at an angle of attack (referred to as AoA hereafter) of $30^\circ$. A towing system was employed to generate impulsively started linear motion for the flat plate. A stainless-steel flat plate with chord length of 2.54 cm, thickness of 0.08 cm, and aspect ratio of 5 was towed in the same plexiglass tank described in section \ref{sec:setup vortex ring}. The towing system consisted of a ball screw rail (Stepperonline) driven by a stepper motor (Stepperonline DM542T) at a constant speed. The motor was controlled by an Arduino board. The flat plate was towed at a Reynolds number of 750. The detailed experimental procedure was reported in our previous work \cite{sun2025vortex}.

% The surrounding fluid was seeded with silver-coated glass spheres (Potters Industries LLC) with medium diameter of 10 um. The seeding particles were illuminated by a laser sheet generated from a 2W 532 nm continuous laser (Ultralasers Inc). The laser sheet was generated by the laser beam passing through a plano-concave lens and a bi-convex lens. The laser sheet was approximately 1 mm and was aligned with the mid-span of the flat plate.

% The images of the flow field were collected through a high-speed camera (Krontech Chronos 2.1) with a 100 mm f/2.8 lens (Venus Optics Laowa). The camera was consistently operated at 1000 frames per second with a fixed exposure time of 250 µs.

\subsection{Turbulent boundary layer}
The experiments were conducted in the high-speed low-turbulence water flume at the Iowa Institute of Hydraulic Research (IIHR) Laboratory at the University of Iowa. The water flume has a test section of 8.2 m long, 0.6 m wide, and 0.4 m deep, with a maximum bulk flow velocity of 1.6 m/s. The turbulent boundary layer (TBL) was developed over a long flat acrylic plate with an elliptic leading edge measuring 1.98 m in length and 0.46 m in width. The flat plate was installed 0.1 m below the free surface to minimize disturbances of unsteady free surfaces. The trailing end of the flat plate is equipped with a flap with sharp end edge that has freely adjustable pitching angle. The flap angle was set at around 15$^circ$ to eliminate the pressure gradients in TBLs. A 1 mm thick, 4 mm wide trip wire was installed at the leading edge of the flat plate to trigger the development of turbulence. The TBL was measured at 1.2 m downstream from the leading edge. The long developing distance ensures fully developed TBLs. 

Here, we are showing representative TBL case with a bulk velocity of 0.38 m/s and a Reynolds number ($Re_\theta$) $\sim$ 1800, where $\theta$ is the momentum thickness. The TBL was seeded with neutrally buoyant silver-coated particles ($\sim13$ $\mu$m diameter, Potters Industries LLC) and was illuminated by 2 mm thick laser sheet (532 nm, 10W, Laserglow). A high-speed camera (IDT, XSM 3520 mini) equipped with a 50 mm canon lens was used to record raw particle images at 800 frame per second over an extended period of 15 seconds. The detailed experimental procedure was reported in our previous work \cite{Wang_Gharib_2020}.

\section{Post-processing exposure control}
Traditionally, the exposure time is controlled by the duration of the film/sensor exposure or by controlling the duration of the light with a continuously exposed sensor. Stacking the particle images allows for the control of exposure time during post-processing by changing the number of images to be stacked. 

The same flow field can appear different depending on the duration of the exposure. This is especially true for unsteady flow. A short exposure time can give nothing but a speckle image with minimal motion blur. A long exposure time can muddle the flow feature with the overlapping of many particle paths. Because different regions of the flow field can have different flow speed, certain regions may appear as speckles, and other regions appear overexposed. Post-experiment control of exposure time allows one to experiment with time integration length to provide the most complete picture of the flow. The length of time integration can be extended to reveal the flow structure of the slow region and shortened to reveal the fast flow region.

\subsection{Vortex ring}
\begin{figure}
    \centering
    \includegraphics[width=1\linewidth]{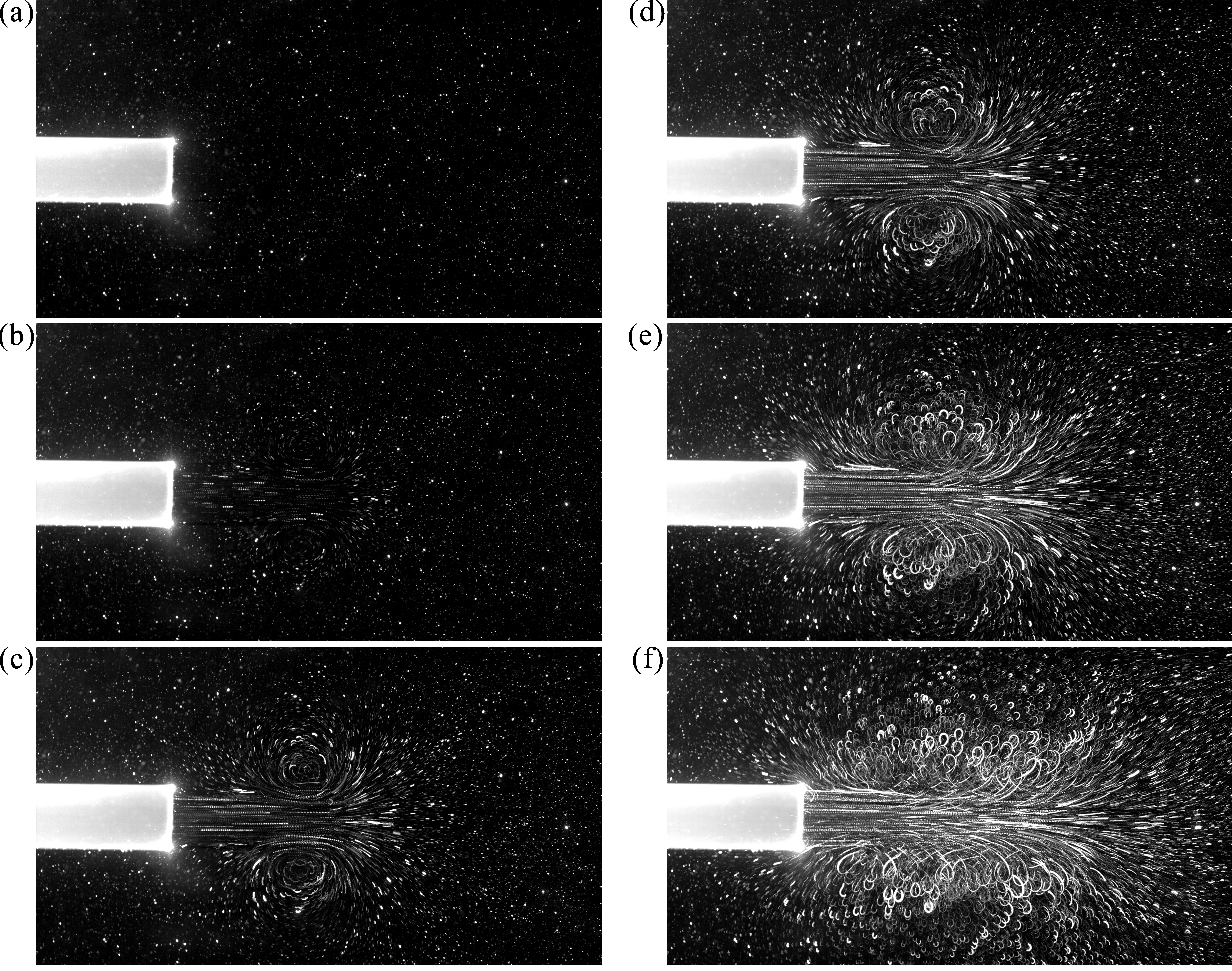}
    \caption{Vortex ring exposure time control. a) Single image, b) 5-Image stack, c) 25-Image stack, d) 50-Image stack, e) 100-Image stack, f) 200-Image stack. The presence of vortex begins to be visible around 25-image stacking. However, the long integration of particle paths leads to different images and highlights the path of individual particles that loop like a pig tail.}
    \label{fig:vortex ring exposure}
\end{figure}

The vortex ring visualized with different exposure times is shown in Fig. \ref{fig:vortex ring exposure}. While the 5-image stack reveals the vortex ring (Fig. \ref{fig:vortex ring exposure}(b)), the presence of the vortical flow is much clearer in the 25-image stack (Fig. \ref{fig:vortex ring exposure}(c)). With clearer visualization, features of the vortex ring such as its core and symmetry \cite{didden1979}, can be easily observed. With a longer exposure time, different characteristics of vortex ring flow are revealed. Instead of a coherent vortex ring, looped trajectories of particles is the most prominent feature. This shows that fluid is initially pushed out and pulled back towards the vortex path as the vortex ring passes by.

\subsection{Impulsively started airfoil at $30^\circ$ AoA}
Here, the growing LEV on top of the airfoil as well as the starting vortex shed from the trailing edge are clearly visible. However, the two vortices have different flow speed. With the longer integration, the trailing edge vortex, whose flow is slower, is made clearer without much overlap between particle paths. However, the LEV is most clearly seen without the overlapping particle paths at 100 image overlap.
\begin{figure}
    \centering
    \includegraphics[width=1\linewidth]{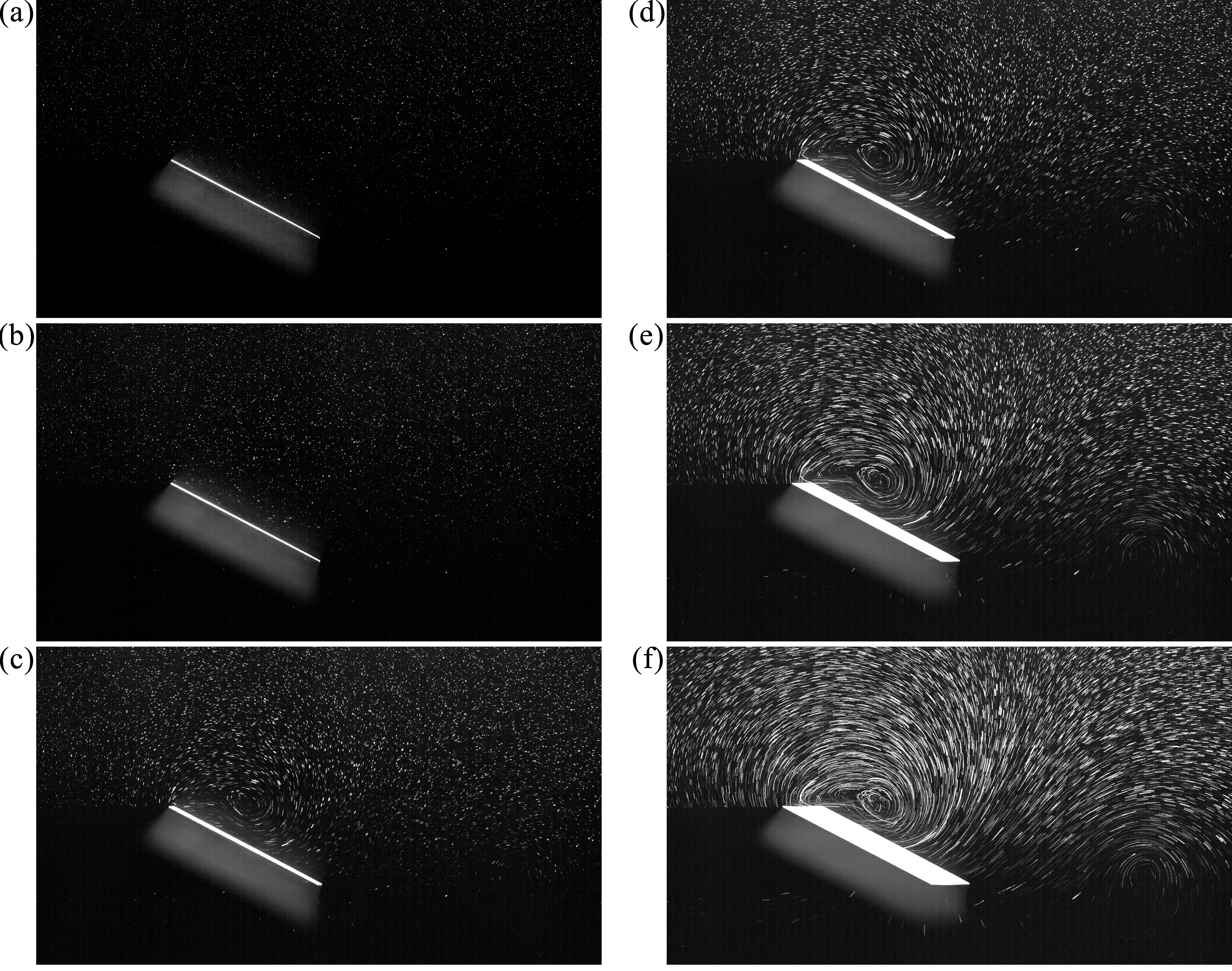}
    \caption{Impulsively started airfoil ($30^\circ$ AoA) exposure time control. a) Single image, b) 5-image stack, C) 25-image stack, D) 50-image stack, e) 100-images stack, f) 200-images stack. Here, too, the presence of LEV begins to be visible around 25-image stacking. Top of the trailing edge vortex is also visible at the bottom right corner in the 50-image stack.}
    \label{fig:lev exposure}
\end{figure}

\subsection{Turbulent boundary layer}
\begin{figure}
    \centering
    \includegraphics[width=1\linewidth]{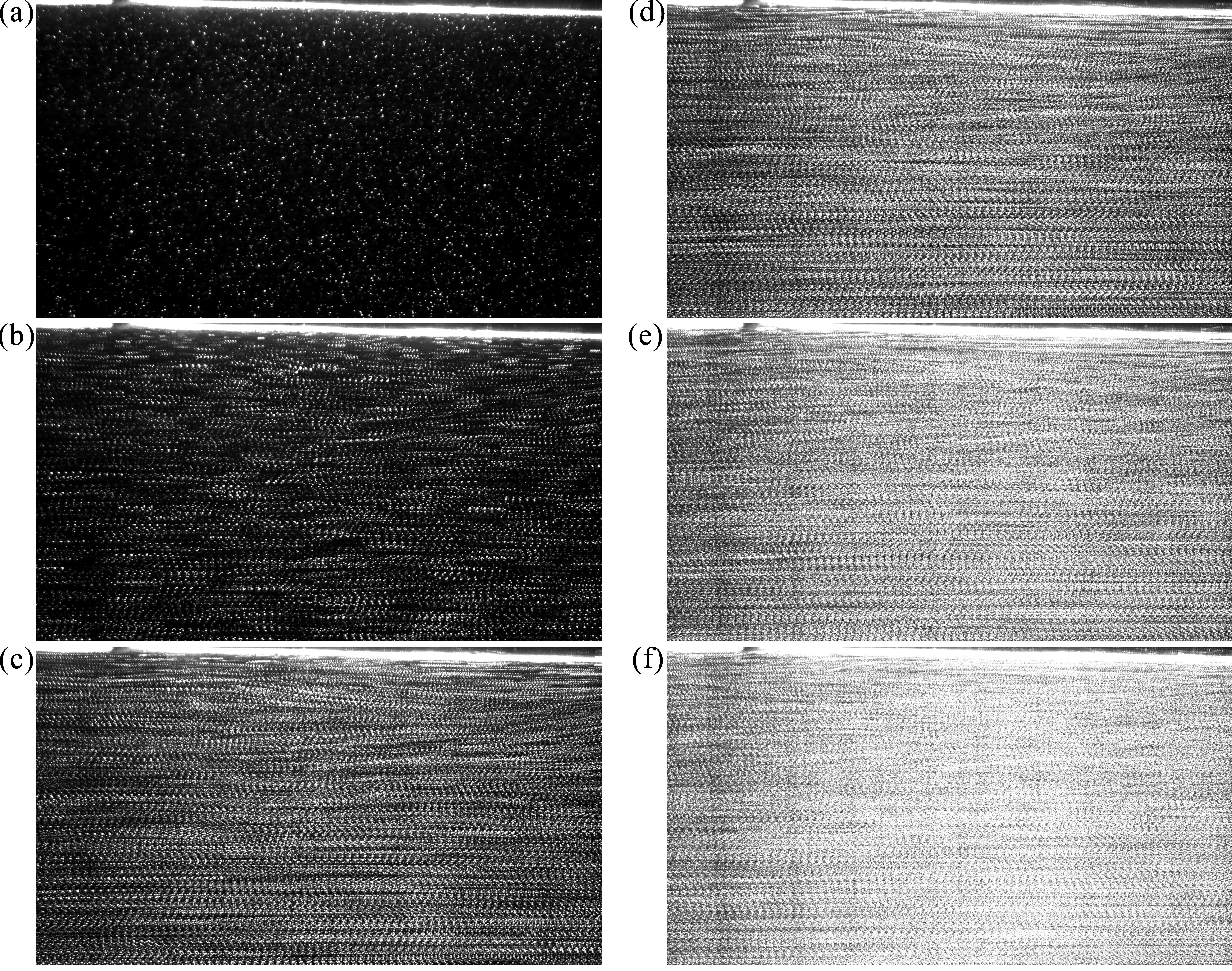}
    \caption{Turbulent boundary layer (TBL) exposure time control. a) Single image, b) 5-image stack, c) 25-image stack, d) 50-image stack, e) 100-image stack, f) 200-image stack. The most informative visualization here is the 5-image stack, showing variation in flow speed due to the turbulent boundary layer. Near the wall, the pathlines are short, and away from the wall, much longer pathlines are observed. Some fluctuation inherent to the TBL can be gleaned from non-parallel pathlines.}
    \label{fig:tbl exposure}
\end{figure}

As expected, the most noticeable feature is that the flow near the wall is slower (Fig. \ref{fig:tbl exposure}). However, the fast mean flow of the TBL hides a relatively smaller fluctuating component. Thus, the flow does not show any other flow features such as a cascade of vortices.
As the number of images in the stack increases, it is harder to identify the slower flow near the wall, due to the overlapping particle trajectories. 

\section{Encoding time}
\subsection{Time-lapsed color coded multi-pathline}
\begin{figure}
    \centering
    \includegraphics[width=0.5\linewidth]{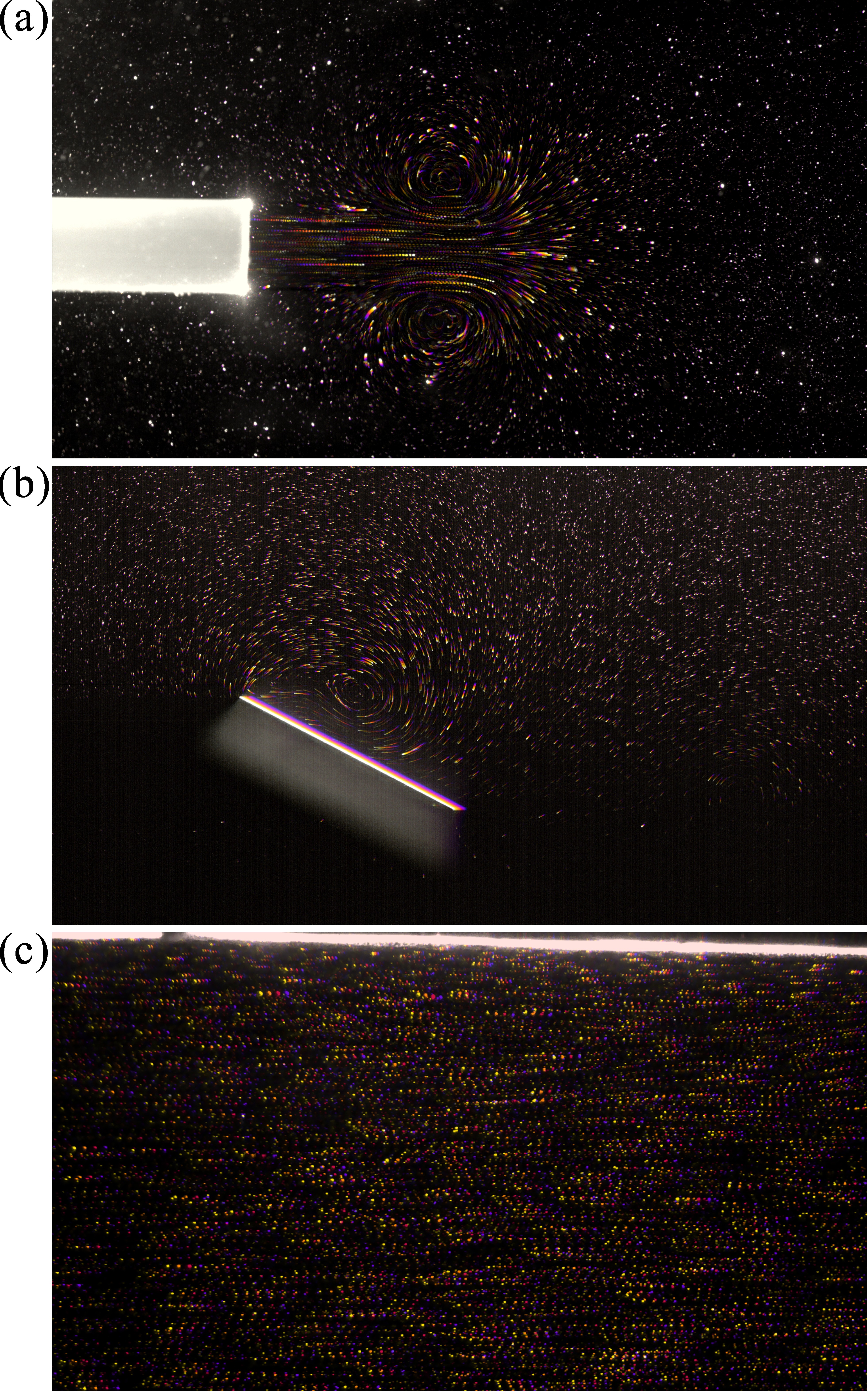}
    \caption{Color-coded multi-pathline. a) Vortex ring. 50-Image stack. b) Leading edge vortex. 50-Image stack. c) Turbulent boundary layer. 5-Image stack. In all images, the direction of the flow is encoded by the color gradient.}
    \label{fig:color timelapse}
\end{figure}

Another advantage of digitally stacking images is that it can also incorporate temporal information. Each image can be assigned a color corresponding to the time prior to stacking. A colormap with gradual changes in color gradient is particularly useful in showing a smooth particle path. In Fig. \ref{fig:color timelapse}, the color-coded multi-pathline was used for visualizing the three flow cases. From purple to yellow, the temporal variation in particle trajectory is encoded using “Fire” colormap LUT (look-up table, predetermined array of colormap). When selecting a colormap, we recommend a sequential LUT, rather than a diverging LUT. In addition to showing the major fluid dynamical features, the detailed flow direction within the structure is visualized.

\section{Changing frame of reference}
Changing the frame of reference provides an opportunity to highlight different flow features. The multi-pathline visualization appears different depending on the frame of reference.  This is because pathline visualization is, by nature, a Lagrangian representation of the flow, and the particle paths are not Galilean invariant:
\begin{equation}
    \underline{x}_0 + \int_0^t \underline{u}(\underline{x},t')\,dt' \neq \underline{x}_0 + \int_0^t \left[\underline{u}(\underline{x},t')-u_\textrm{ref}\right]\,dt',
\end{equation}
where $u_\textrm{ref}$ is the speed of the reference frame. Thus, an ability to change frame of reference can enhance the versatility of the multi-pathline visualization method.

While the frame of reference can be changed experimentally by moving the camera at a certain speed while recording, it can also be achieved through post-processing high-speed camera footage. By shifting the images relative to the previous image at a desired frame of reference speed, one can effectively change the frame of reference. This also means that towing experiments can be turned into water tunnel experiments and vice versa.

Adjusting the frame of reference will often require shifting the image at a subpixel resolution. The simplest way to achieve subpixel shifting of the image is through a bilinear interpolation. Bilinear Interpolation is an average of neighboring pixels weighted by the amount of subpixel shift. If the pixel is shifted from $(x_o,y_o)$ location by $\Delta x=[0,1)$ in the $x$-direction and by $\Delta y=[0,1)$ in the $y$-direction, then the shifted pixel intensity, $I$, is given by
\begin{eqnarray}
    I_\textrm{shifted}(x_o+\Delta x,y_o+\Delta y) \nonumber\\
    && =I(x_o,y_o)\cdot(1-\Delta x)\cdot(1-\Delta y) + I(x_o+1,y_o)\cdot\Delta x\cdot(1-\Delta y) \nonumber\\
    && +I(x_o,y_o+1)\cdot\Delta y\cdot(1-\Delta x) + I(x_o+1,y_o+1)\cdot\Delta x\cdot\Delta y.
\end{eqnarray}

While in most cases, a useful frame of reference is known a priori, other flow-dependent frames of reference, such as speed of vortex propagation, may be known only after the experiment. The post-processing implementation of the frame of reference change allows quick sweep of reference frame changes, helping identifying relevant $u_\textrm{ref}$ through many different trials. Moreover, the speed of the salient flow feature or the object of interest can be non-uniform. The change of reference frame during the post-processing allows for easier implementation of complex frame of reference shifting, such as non-straight path and accelerating frame.

Combined with frame-independent quantitative field data such as strain field or vorticity fields, these visualizations can enhance our perception of the flow phenomena. This will be demonstrated in the following three examples.

\subsection{Vortex ring}
\begin{figure}
    \centering
    \includegraphics[width=0.5\linewidth]{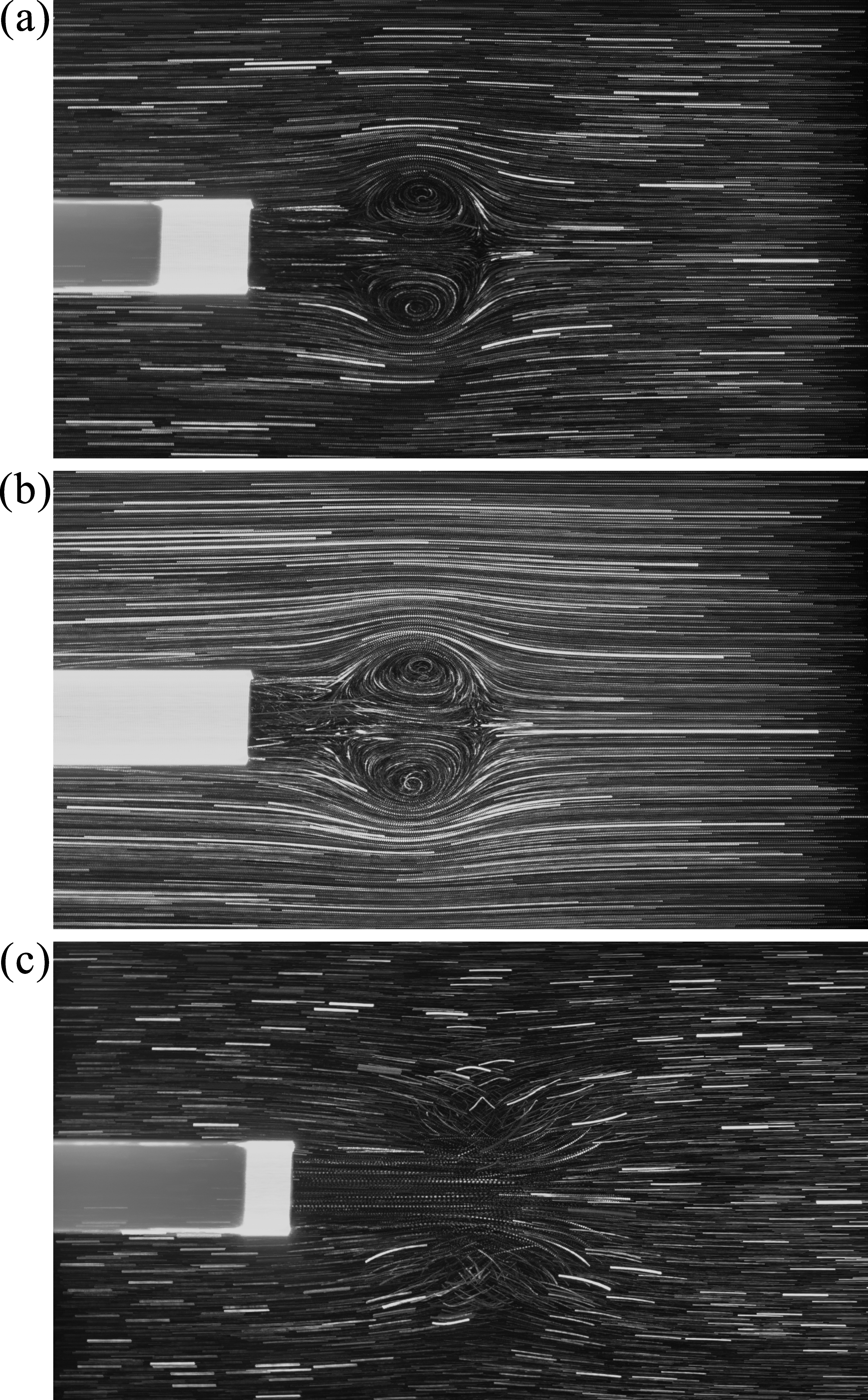}
    \caption{Frame shifted : Vortex ring. a) Frame shifted, at the speed of the vortex ring. 50-Image stack. b) Frame shifted, at the speed of the vortex ring. 200-Image stack. c) Frame shifted, at 1.5 times the speed of the vortex ring. 50-Image stack.}
    \label{fig:shift vortex ring}
\end{figure}

Compared to the vortex ring in a stationary frame (Fig. \ref{fig:vortex ring exposure}(c)), the same flow in a propagating vortex frame shows similar but distinct features even for a small integration time (50 frames, Fig. \ref{fig:shift vortex ring}(a))). Notably, in the frame of the vortex ring, the stagnation point in the front is noticeable and the streamline that encapsulates the vortex is readily visible. For a longer integration time, the difference in visualization between the stationary and shifted frame of reference is stark (200 frames, Fig. \ref{fig:vortex ring exposure} (f) vs Fig. \ref{fig:shift vortex ring}(b)). In the frame of propagating vortex, the vortical structure maintains its coherence. Thus, other than the brightness, only minimal difference is seen between the short and long integration visualization in the vortex frame.

When the frame is shifted so that the frame moves faster than the vortex ring, the vortex ring’s swirly character is lost. Instead, the flow caves towards the vortex core (Fig. \ref{fig:shift vortex ring}(c)).

\subsection{Impulsively started airfoil at $30^\circ$ AoA}
\begin{figure}
    \centering
    \includegraphics[width=0.5\linewidth]{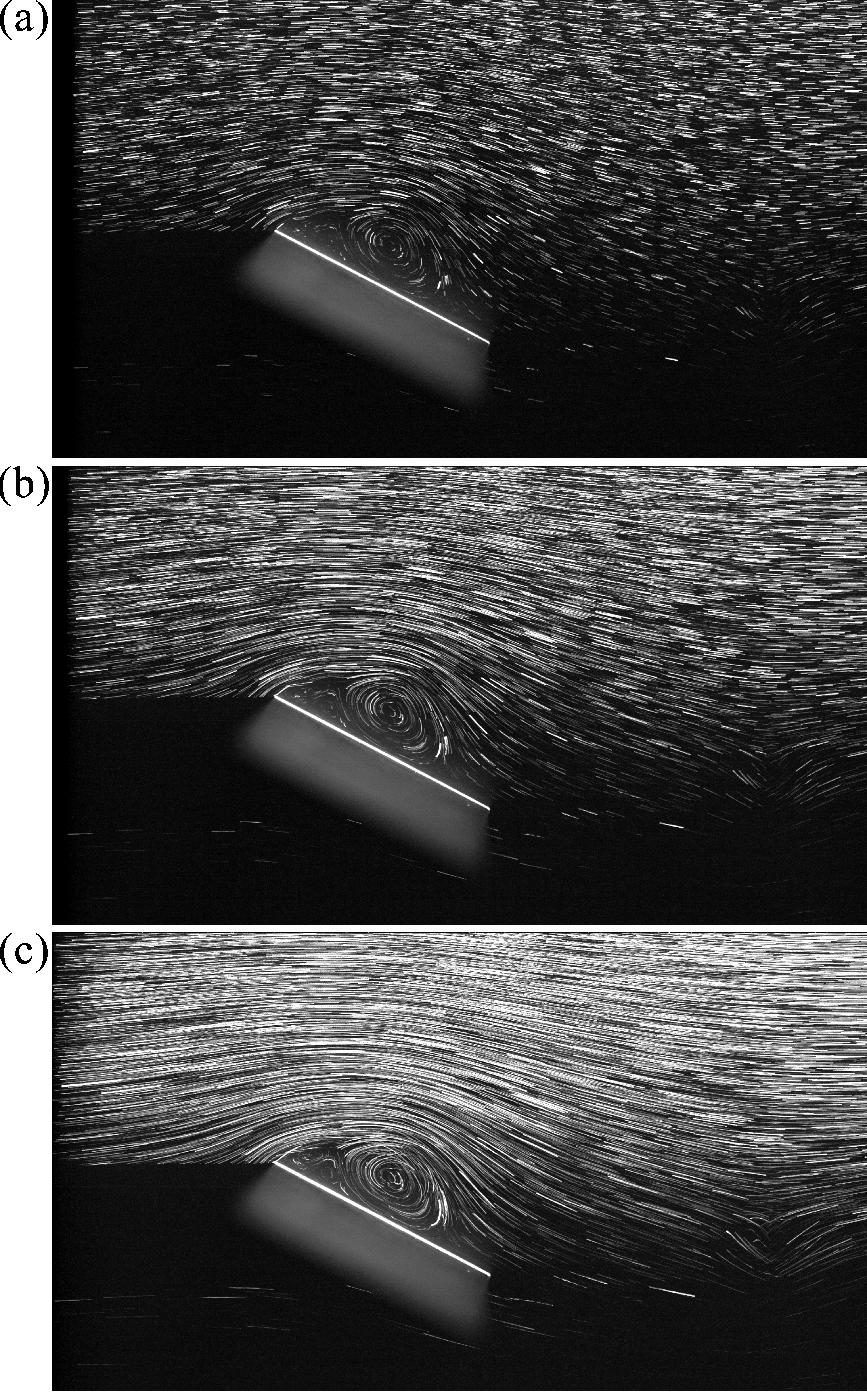}
    \caption{Frame shifted: Impulsively started inclined plate at $30^\circ$ AoA. a-c) Frame shifted to the frame of the airfoil. a) 50-Image stack. b) 100-Image stack. c) 200-Image stack.}
    \label{fig:shift lev}
\end{figure}

In the frame of the airfoil, the LEV appears as a different shape. Similar to the vortex ring, the streamline wraps around the growing vortex, delineating the vortex from the outer region free of vorticity. Moreover, the longer time-integrated visualization shows two counter-rotating recirculation zones near the leading edge (Fig. \ref{fig:shift lev}(c-d)), which is not depicted in the stationary reference frame shown in Fig. \ref{fig:lev exposure}. On the other hand, the trailing edge vortex that is in the frame of the freestream velocity is harder to identify in the frame of the airfoil. Similar to the vortex ring in the frame moving faster than the vortex ring, only a slight kink in the flow is visible where the trailing edge vortex was formed. Lastly, the Kutta condition satisfied at the trailing edge is only visible in the shifted frame of reference (compare Fig. \ref{fig:lev exposure} vs. Fig. \ref{fig:shift lev}).

\subsection{Turbulent boundary layer}
\begin{figure}
    \centering
    \includegraphics[width=0.5\linewidth]{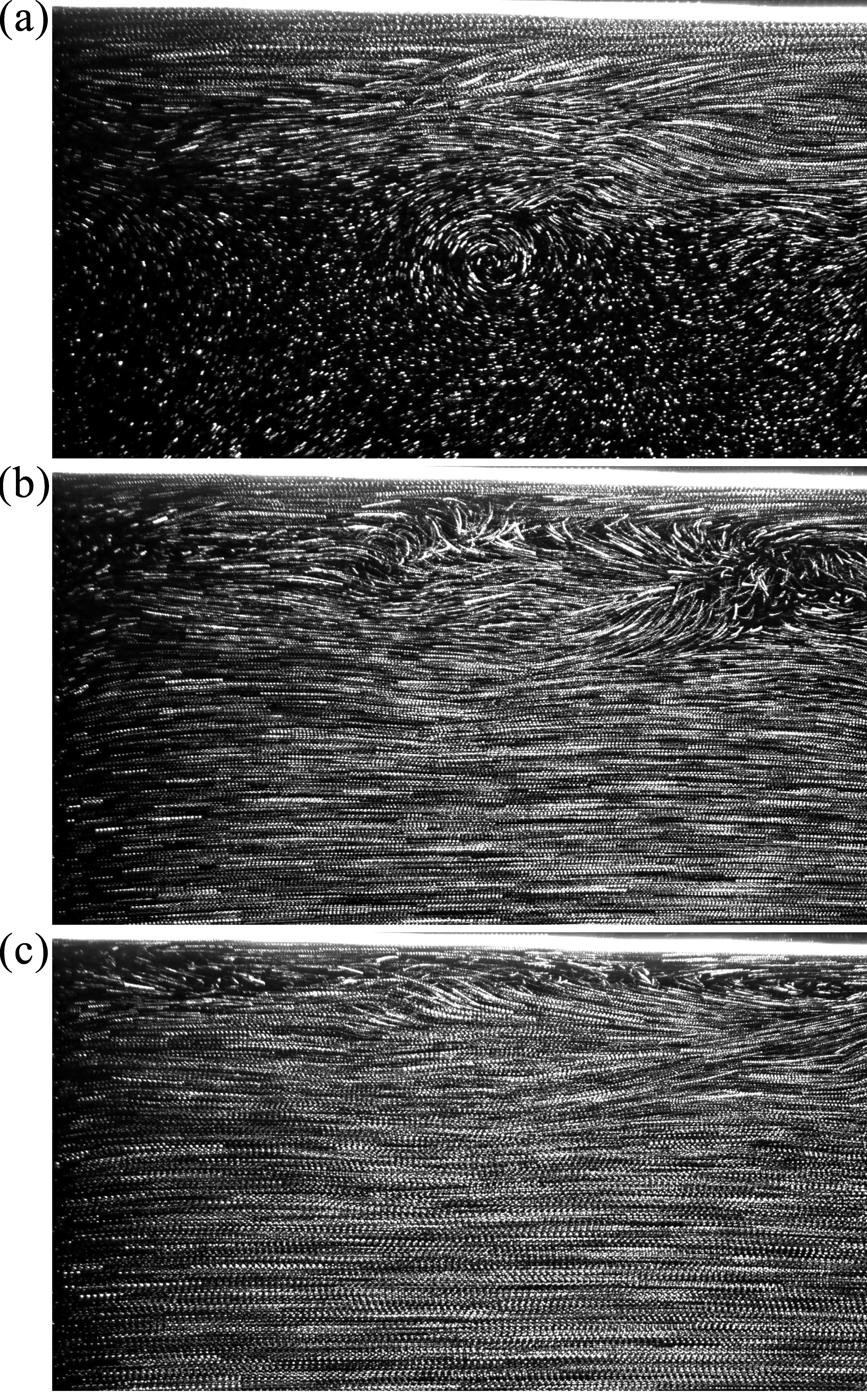}
    \caption{Frame shifted: Turbulent boundary layer. a) Frame shifted, mean flow speed. b) Frame shifted, 50$\%$ of the mean flow speed. c) Frame shifted, 90$\%$ of the mean flow speed. (a-c) are all 5-image stacks.}
    \label{fig:shift tbl}
\end{figure}

In a fixed frame, the turbulent boundary layer appears as a somewhat featureless parallel flow. Frame shifting can visualize the swirls within the turbulent boundary layer. Subtracting the mean flow reveals the vortex at the outer edge of the turbulent boundary layer (Fig. \ref{fig:shift tbl}(a)). Reduced speed of the reference frame reveals unsteady swirl at the center of the boundary layer (Fig. \ref{fig:shift tbl}(b)), and further reduction shows the swirl near the solid boundary (Fig. \ref{fig:shift tbl}(c)).  

\section{Conclusion}
The ability to see the flow deeply influences our understanding of fluid dynamics. The multi-pathline visualization provides a quick way of seeing the flow phenomena under different perspectives, which will lead to creative insight, sharpen our intuition, and bring out the innately aesthetic features of the flows based on existing raw experimental data. While this paper demonstrates stacking images into a snapshot, a continuous sequence of multi-pathline images can be made into a video for enhanced visualization of the flow field.

We hope that the simplicity of the post-processing tools presented here, which do not require additional experiments, will empower fluid dynamicists to employ the algorithm to the flow field of their interest. 

\begin{acknowledgments}
We gratefully acknowledge the following supports: Y.S. and C.R. are supported by NSF CMMI-2042740 and NSF CBET-2442036; C.W. is support by ONR N00014-22-1-2233; E.J. is supported by Colman Family Endowed Fellowship at Cornell University; F.F. is supported by Engineering Learning Initiatives at Cornell University.
\end{acknowledgments}

\appendix
% \section{Appendix}

\bibliography{references}% Produces the bibliography via BibTeX.

\end{document}